\documentclass[11pt,a4paper]{article}

\usepackage[utf8x]{inputenc}
\usepackage[T1]{fontenc}
\usepackage{color}
\usepackage{amssymb}
\usepackage{amsmath,amsfonts}
\usepackage{graphicx}
\usepackage{mathtools}
\usepackage{ragged2e}
\usepackage{psfrag}
\usepackage{relsize}

\usepackage{bm}
\usepackage{mathrsfs}
\usepackage{graphicx}
\usepackage{verbatim} 
\usepackage[english]{babel}
\usepackage{hyperref}
\hypersetup{
citecolor=red,
colorlinks=true,
filecolor=red,
linkcolor=blue,
linktocpage=true,
urlcolor=blue
} 
\usepackage[titletoc,toc,title]{appendix}
\numberwithin{equation}{section}
\usepackage{cleveref}
\usepackage{cite}
\usepackage{epsfig}
\usepackage{float}
\usepackage{enumitem}
\usepackage[font={footnotesize,it}]{caption}
\usepackage{cases}
\usepackage{amssymb,tikz,hyperref,empheq}

\newcommand\trick[1]{}
\newcommand{\be}{\begin{equation}} 
\newcommand{\ee}{\end{equation}}
\newcommand{\eq}[1]{(\ref{#1})}
\newcommand{\bit}{\begin{itemize}}  \newcommand{\eit}{\end{itemize}}
\newcommand{\ben}{\begin{enumerate}}  \newcommand{\een}{\end{enumerate}}

\newcommand{\rf}[1]{(\ref{#1})}

\def\bd{\begin{document}}
\def\ed{\end{document}}
\def\bea{\begin{eqnarray}}
\def\eea{\end{eqnarray}}
\let\bm=\bibitem

\def\la{\langle}
\def\ra{\rangle}

\def\npb#1#2#3{Nucl. Phys. {\bf{B#1}} #3 (#2)}
\def\plb#1#2#3{Phys. Lett. {\bf{#1B}} #3 (#2)}
\def\prl#1#2#3{Phys. Rev. Lett. {\bf{#1}} #3 (#2)}
\def\prd#1#2#3{Phys. Rev. {D bf{#1}} #3 (#2)}
\def\cmp#1#2#3{Comm. Math. Phys. {\bf{#1}} #3 (#2)}
\def\cqg#1#2#3{Class. Quantum Grav. {\bf{#1}} #3 (#2)}
\def\nppsa#1#2#3{Nucl. Phys. B (Proc. Suppl.) {\bf{#1A}}#3 (#2)}
\def\ap#1#2#3{Ann. of Phys. {\bf{#1}} #3 (#2)}
\def\ijmp#1#2#3{Int. J. Mod. Phys. {\bf{A#1}} #3 (#2)}
\def\rmp#1#2#3{Rev. Mod. Phys. {\bf{#1}} #3 (#2)}
\def\mpla#1#2#3{Mod. Phys. Lett. {\bf A#1} #3 (#2)}
\def\jhep#1#2#3{J. High Energy Phys. {\bf #1} #3 (#2)}
\def\atmp#1#2#3{Adv. Theor. Math. Phys. {\bf #1} #3 (#2)}

\def\sst{\scriptscriptstyle}
\def\thetabar{\bar\theta}
\def\Tr{{\rm Tr}}
\def\XN{X^{(N)}}
\def\XNp{X^{(N')}}
\def\An{A^{(n)}}

\def\a{\alpha}      \def\da{{\dot\alpha}}  \def\dA{{\dot A}}
\def\b{\beta}       \def\db{{\dot\beta}}
\def\g{\gamma}  \def\G{\Gamma}  \def\dc{{\dot\gamma}}
\def\d{\delta}  \def\D{\Delta}  \def\ddt{\dot\delta}
\def\e{\epsilon}
\def\ve{\varepsilon}
\def\uve{\upvarepsilon}
\def\f{\phi}    \def\F{\Phi}    \def\vvf{\f}
\def\vphi{\varphi}
\def\h{\eta}
\def\k{\kappa}
\def\l{\lambda} \def\L{\Lambda}
\def\m{\mu} \def\n{\nu}
\def\o{\omega}
\def\p{\pi} \def\P{\Pi}
\def\r{\rho}
\def\s{\sigma}  \def\S{\Sigma}
\def\t{\tau}
\def\th{\theta} \def\Th{\Theta} \def\vth{\vartheta}
\def\X{\Xeta}
\def\z{\zeta}

\def\na{\nabla}

\def\cA{{\cal A}} \def\cB{{\cal B}} \def\cC{{\cal C}}
\def\cD{{\cal D}} \def\cE{{\cal E}} \def\cF{{\cal F}}
\def\cG{{\cal G}} \def\cH{{\cal H}} \def\cI{{\cal I}}
\def\cJ{{\mathscr J}} \def\cK{{\cal K}} \def\cL{{\cal L}}
\def\cM{{\cal M}} \def\cN{{\cal N}} \def\cO{{\cal O}}
\def\cP{{\cal P}} \def\cQ{{\cal Q}} \def\cR{{\cal R}}
\def\cS{{\cal S}} \def\cT{{\cal T}} \def\cU{{\cal U}}
\def\cV{{\cal V}} \def\cW{{\cal W}} \def\cX{{\cal X}}
\def\cY{{\cal Y}} \def\cZ{{\cal Z}}
\def\ct{{\cal t}}

\def\ua{\underline{\alpha}}
\def\uc{\underline{\phantom{\alpha}}\!\!\!\gamma}
\def\um{\underline{\mu}}
\def\ud{\underline\delta}
\def\ue{\underline\epsilon}
\def\una{\underline a}\def\unA{\underline A}
\def\unb{\underline b}\def\unB{\underline B}
\def\unc{\underline c}\def\unC{\underline C}
\def\und{\underline d}\def\unD{\underline D}
\def\une{\underline e}\def\unE{\underline E}
\def\unf{\underline{\phantom{e}}\!\!\!\! f}\def\unF{\underline F}
\def\unm{\underline m}\def\unM{{\underline M}}
\def\unn{\underline n}\def\unN{{\underline N}}
\def\unp{\underline{\phantom{a}}\!\!\! p}\def\unP{\underline P}
\def\unq{\underline{\phantom{a}}\!\!\! q}
\def\unQ{\underline{\phantom{A}}\!\!\!\! Q}
\def\unH{\underline{H}}

\def\As {{A \hspace{-6.4pt} \slash}\;}
\def\bs {{b \hspace{-6.4pt} \slash}\;}
\def\Ds {{D \hspace{-6.4pt} \slash}\;}
\def\Gts {{\Gt \hspace{-6.4pt} \slash}\;}
\def\ds {{\del \hspace{-6.4pt} \slash}\;}
\def\ss {{\s \hspace{-6.4pt} \slash}\;}
\def\ks {{ k \hspace{-6.4pt} \slash}\;}
\def\ps {{p \hspace{-6.4pt} \slash}\;}
\def\xs {{x \hspace{-6.4pt} \slash}\;}
\def\pas {{{p_1} \hspace{-6.4pt} \slash}\;}
\def\pbs {{{p_2} \hspace{-6.4pt} \slash}\;}
\def\cFs {{{\cal F} \hspace{-6.4pt} \slash}\;}
\def\Dss {{D \hspace{-7.5pt} \slash}\;}
\def\dss {{\del \hspace{-7.0pt} \slash}\;}

\def\Ah{{\hat{A}}}
\def\Dh{{\hat{D}}}
\def\Gh{{\hat{G}}}
\def\Fh{{\hat{F}}}
\def\Ih{{\hat{I}}}
\def\Jh{{\hat{J}}}
\def\Kh{{\hat{K}}}
\def\Lh{{\hat{L}}}
\def\Ph{{\hat{P}}}
\def\Rh{{\hat{R}}}
\def\Vh{{\hat{V}}}
\def\Xh{{\hat{X}}}

\def\ah{{\hat{\a}}}
\def\bh{{\hat{\b}}}
\def\gh{{\hat{\g}}}
\def\dh{{\hat{\d}}}
\def\rh{{\hat{\r}}}
\def\hh{\hat{h}}
\def\uh{\hat{u}}
\def\xh{\hat{x}}
\def\yh{\hat{y}}
\def\ph{\hat{p}}
\def\xih{\hat{\xi}}
\def\chih{\hat{\chi}}
\def\Psih{\hat{\Psi}}
\def\phih{\hat{\phi}}

\def\psit{\tilde{\psi}}
\def\Psit{\tilde{\Psi}}
\def\Psibt{\tilde{\bar{Psi}}}

\def\lambdat{\tilde {\lambda}}
\def\st{\tilde{\sigma}}

\def\delt{\tilde{\delta}}
\def\Phit{\tilde{\Phi}}
\def\Phitb{\overline{\tilde{Phi}}}
\def\tht{\tilde{\th}}
\def\lt{\tilde{\l}}
\def\chit{\tilde{\chi}}
\def\phit{\tilde{\phi}}

\def\At{\tilde{A}}
\def\Bt{\tilde{B}}
\def\Ct{\tilde{C}}
\def\Dt{\tilde{D}}
\def\Et{\tilde{E}}
\def\Ft{\tilde{F}}
\def\Gt{\tilde{G}}
\def\Ht{\tilde{H}}
\def\It{\tilde{I}}
\def\Jt{\tilde{J}}
\def\Pt{\tilde{P}}
\def\Ot{\tilde{O}}
\def\Mt{\tilde{M }}
\def\Nt{\tilde{N}}
\def\St{\tilde{S}}
\def\Vt{\tilde{V}}
\def\Xt{\tilde{X}}
\def\at{\tilde{a}}
\def\ct{\tilde{c}}
\def\dt{\tilde{d}}
\def\htt{\tilde{h}}
\def\ft{\tilde{f}}
\def\gt{\tilde{\gamma}}
\def\pt{\tilde{p}}
\def\qt{\tilde{q}}
\def\rt{\tilde{r}}
\def\nt{\tilde{n}}
\def\ut{\tilde{u}}
\def\wt{\tilde{w}}
\def\zt{\tilde{z}}
\def\xt{\tilde{x}}
\def\yt{\tilde{y}}
\def\Psit{\tilde{\Psi}}
\def\phit{\tilde{\phi}}
\def\tD{\tilde{\D}}

\def\eb{\bar{\epsilon}}
\def\delb{\bar{\partial}}
\def\thb{\bar{\theta}}
\def\mub{\bar{\mu}}
\def\lamb{\bar{\l}}
\def\psib{\bar{\psi}}
\def\sb{\bar{\sigma}}
\def\xib{\bar{\xi}}
\def\chib{\bar{\chi}}

\def\Psib{\bar{\Psi}}
\def\Phib{\bar{\Phi}}
\def\Lamb{\bar{\Lambda}}
\def\Sb{{\overline \Sigma}}
\def\cb{\bar{c}}
\def\hb{\bar{h}}
\def\qb{\bar{q}}
\def\wb{\bar{w}}
\def\ub{\bar{u}}
\def\zb{{\bar{z}}}
\def\Hb{\bar{H}}
\def\Qb{{\bar Q}}
\def\Omegab{\overline{\Omega}}
\def\ob{\overline{\omega}}

\def\Ab{{\overline A}} \def\Bb{{\overline B}} \def\Cb{{\overline C}}
\def\Db{{\overline D}} \def\Eb{{\overline E}} \def\Fb{{\overline F}}
\def\Gb{{\overline G}}
\def\Ib{{\overline I}}
\def\Jb{{\overline J}} \def\Kb{{\overline K}} \def\Lb{{\overline L}}
\def\Mb{{\overline M}} \def\Nb{{\overline N}} \def\Ob{{\overline O}}
\def\Pb{{\overline P}}  \def\Rb{{\overline R}}
 \def\Tb{{\overline T}} \def\Ub{{\overline U}}
\def\Vb{{\overline V}} \def\Wb{{\overline W}} \def\Xb{{\overline X}}
\def\Yb{{\overline Y}} \def\Zb{{\overline Z}}

\def\fb{{\overline f}}
\def\gb{{\overline g}}
\def\nb{{\overline n}}
\def\mb{{\overline m}}
\def\lb{{\overline l}}
\def\yb{{\overline y}}

\def\ldel{{\overleftarrow{\del}}}
\def\rdel{{\overrightarrow{\del}}}
\def\ldeldel{{\overleftarrow{\del^2}}}
\def\rdeldel{{\overrightarrow{\del^2}}}
\def\ldelb{{\overleftarrow{\bar{\del}}}}
\def\rdelb{{\overrightarrow{\bar{\del}}}}

\def\ba{{\bf a}}
\def\bk{{\bf k}}
\def\bl{{\bf l}}
\def\bp{{\bf p}}
\def\bq{{\bf q}}

\def\br{{\bf r}}
\def\bt{{\bf t}}
\def\bu{{\bf u}}
\def\bv{{\bf v}}
\def\bx{{\bf x}}
\def\by{{\bf y}}
\def\bA{{\bf A}}
\def\bR{{\bf R}}
\def\bV{{\bf V}}

\def\bz{{\boldsymbol{\zeta}}}

\def\bone{{\bf 1}}

\def\va{{\vec a}}
\def\vk{{\vec k}}
\def\vp{{\vec p}}
\def\vq{{\vec q}}
\def\vx{{\vec x}}
\def\vy{{\vec y}}
\def\vu{{\vec u}}
\def\vv{{\vec v}}
\def \vH{{\vec H}}
\def \vg{{\vec g}}

\def\vs{{\vec \sigma}}
\def\vtau{{\vec \tau}}

\newcommand{\ov}[1]{\overrightarrow{#1}}

\def\frA{\mathfrak{A}}
\def\frB{\mathfrak{B}}
\def\frC{\mathfrak{C}}
\def\frD{\mathfrak{D}}
\def\frE{\mathfrak{E}}
\def\frF{\mathfrak{F}}
\def\frG{\mathfrak{G}}
\def\frH{\mathfrak{H}}
\def\frM{\mathfrak{M}}
\def\frN{\mathfrak{N}}
\def\frR{\mathfrak{R}}
\def\frW{\mathfrak{W}}

\def\fra{\mathfrak{a}}
\def\frb{\mathfrak{b}}
\def\frf{\mathfrak{f}}
\def\frg{\mathfrak{g}}
\def\frh{\mathfrak{h}}
\def\frl{\mathfrak{l}}
\def\frs{\mathfrak{s}}
\def\fri{\mathfrak{i}}
\def\frj{\mathfrak{j}}

\def\ma{\mathfrak{a}}
\def\mg{\mathfrak{g}}
\def\mh{\mathfrak{h}}
\def\mR{\mathfrak{R}}
\def\mN{\mathfrak{N}}

\newcommand{\nn}{{\nonumber}}

\def\d{\delta}\def\D{\Delta}\def\ddt{\dot\delta}

\def\pa{\partial} \def\del{\partial}
\def\xx{\times}
\def\uno{\mbox{1 \kern-.59em {\rm l}}}

\def\trp{^{\top}}
\def\inv{^{-1}}
\def\dag{\dagger}
\def\pr{^{\prime}}

\def\rar{\rightarrow}
\def\lar{\leftarrow}
\def\lrar{\leftrightarrow}

\newcommand{\0}{\,\!}      
\def\one{1\!\!1\,\,}
\def\im{\imath}
\def\jm{\jmath}

\newcommand{\tr}{\mbox{tr}}
\newcommand{\slsh}[1]{/ \!\!\!\! #1}

\newcommand{\1}{\mbox{1}\hspace{-0.25em}\mbox{l}}

\def\vac{|0\rangle}
\def\lvac{\langle 0|}

\def\hlf{\frac{1}{2}}
\def\ove#1{\frac{1}{#1}}
\newcommand{\hot}[1]{\frac{#1}{2}}

\def\Box{\square}
\def\CC {\mathbb{C}}
\def\FF {\mathbb{F}}
\def\RR{\mathbb{R}}
\def\NN{\mathbb{N}}
\def\ZZ{\mathbb{Z}}
\def\bb#1{{\bf #1}}
\def\bcomment#1{}
\def\bfhat#1{{\bf \hat{#1}}}
\def\VEV#1{\left\langle #1\right\rangle}

\newcommand{\ex}[1]{{\rm e}^{#1}} \def\ii{{\rm i}}

\newcommand{\lrbrk}[1]{\left(#1\right)}
\newcommand{\lrsbrk}[1]{\left[#1\right]}
\newcommand{\sfrac}[2]{{\textstyle\frac{#1}{#2}}}

\def\stw{{\sqrt{2}}}

\def\rf {{\rm f}}
\def\ri {{\rm i}}
\def\rj {{\rm j}}
\def\rn {{\rm n}}
\def\rk {{\rm k}}
\def\rl {{\rm l}}
\def\rr {{\rm r}}

\def\rQ {{\scriptscriptstyle \rm \cQ}}
\def\rR {{\scriptscriptstyle \rm \cR}}

\def\cQb{{\cal \Qb}}
\def\cRb{{\cal \Rb}}
\def\cWb{{\cal \Wb}}

\def\fd {{\rm N}}
\def\afd {{\overline{\rm N}}}

\def \II {I\hspace{-.1em}I\hspace{.1em}}
\def \IIA {\mbox{\II A\hspace{.2em}}}
\def \IIB {\mbox{\II B\hspace{.2em}}}
\def \gs {g^s}
\def \ls {\lambda^s}

\def \I {{\cal I}}
\def \qs {q\hspace{-.53em}/\hspace{.15em}}
\def \ks {k\hspace{-.53em}/\hspace{.15em}}
\def \YM {{\mbox{\tiny YM}}}
\def \gym {g_{\YM}}

\def \Lc {\L_c}
\def\IR{\relax{\rm I\kern-.18em R}}
\def \id {{\bf 1}}

\def\cci{\ell}
\def\ccj{\ell'}

\def\bbq{\pmb{q}}
\def\bom{\pmb{\o}}
\def\bJ{\pmb{J}}
\def\bM{\pmb{M}}
\def\bB{\pmb{B}}
\def\bn{\pmb{n}}
\def\bE{\pmb{E}}

\newcommand{\rrr}[1]{\vskip 0.2cm \noindent{\bf #1} ---}

\long\def\symbolfootnote[#1]#2{\begingroup%
\def\thefootnote{\fnsymbol{footnote}}\footnote[#1]{#2}\endgroup}
\long\def\RemarkBox#1{\begin{flushleft}\fbox{\begin{minipage}
{17.5cm}{\bf Remark:} ~#1\end{minipage}}\end{flushleft}}

\newcommand{\nthu}{{\it Department of Physics, National Tsing-Hua University,
    Hsinchu 30013, Taiwan}}

\newcommand{\ctc}{{\it
Center for Theory-Computation-Data Science Research, 
National Tsing-Hua University, Hsinchu 30013, Taiwan}}

\newcommand{\ncts}{{\it Physics Division,
    National Center for Theoretical Sciences, Taipei 10617, Taiwan}}

\begin{document}
\begin{center}
\vspace{20pt}
  
\thispagestyle{empty}
              {\Large \bf  Hawking Radiation from Tunneling in
                \\
                Black Hole Quantum Mechanics}
               
\vspace{25pt}

Chong-Sun Chu

\vspace{0.2cm}              

\vspace{5pt}\nthu\\
\vspace{5pt}\ctc\\
\vspace{5pt}\ncts

\vspace{1cm}

\begin{abstract}
  It was proposed in \cite{Chu:2024qil}
  that a quantum black hole can be described by
  a quantum space configuration of 
  a fuzzy sphere together with a half-filled Fermi sea. In this paper
  we propose that
  the tunneling of the fuzzy sphere system to a small one describes
  the quantum decay of black hole by Hawking radiation.
  Since the Fermi sea
  shrinks and the quantum mechanical Hamiltonian
  conserves fermion number, the amplitude of transition naively
  vanishes unless
  the tunneling path provides exact number of
  zero modes to soak up the excess fermi
  states. We show that a monopole on fuzzy sphere does exactly that.
This fixes the tunneling path. The resulting tunneling rate
reproduces
  Page's result for  the semi-classical decay rate of black hole.
  The
  quantum states released by the monopole corresponds to 
  gravitational Hawking radiation. 
 At the level of  probability, the
 Hawking radiation is found to be given by a Boltzmann distribution  at
 the Hawking temperature.
 One can go beyond the
 probabilistic description by determining the full wave function of the
 multi-partite Hawking quanta. This is possible with a real time 
 formulation of the tunneling process.
 Unitarity is manifest in our quantum mechanics.

\end{abstract}

\end{center}

\newpage
\setcounter{footnote}{0}

\tableofcontents

\section{Introduction}

Black holes pose sharp consistency questions at the interface of gravity,
quantum mechanics, and thermodynamics. It is widely believed that
resolving problems such as providing a microscopic account of
Bekenstein–Hawking entropy
\cite{Bekenstein:1972tm,Bekenstein:1973ur},
understanding the origin of black hole
thermodynamics \cite{Hawking:1975vcx},
and resolving the information paradox
\cite{Hawking:1976ra} posed by Hawking radiation or explaining
the Page curve \cite{Page:1993wv,Page:2013dx} will
provide valuable insights to the construction of the
theory of quantum gravity.


The BFSS matrix quantum
mechanics of supersymmetric D0-branes \cite{Banks:1996vh}  has been
conjectured to provide a non-perturbative definition of the
eleven-dimensional M-theory,
see \cite{Taylor:2001vb,Becker:2006dvp, Lin:2025iir} for review
of the BFSS matrix model.  
The BFSS
and the IKKT \cite{Ishibashi:1996xs}
matrix model have transformed the way we think about spacetime
by replacing it with an effective geometry that emerges from the
many-body quantum dynamics of the underlying matrix degrees of
freedom.
The  idea that quantum gravity can be captured by
the quantum mechanics of a large-$N$ system has also influenced 
the  development of the AdS/CFT duality \cite{Maldacena:1997re}.
Matrix model has  offered valuable
insights into the working of quantum gravity
in flat spacetime,  see for example
\cite{Becker:1997wh,Herderschee:2023bnc,Herderschee:2023pza,Ho:2025htr}.
Nevertheless, achieving a fully explicit account of black hole physics
in the BFSS
or IKKT matrix model has remained an open problem so far.
In particular, it is
unclear how to describe, for example,
a horizon-like geometry, the Bekenstein–Hawking entropy,
or the Hawking radiation
directly from the underlying quantum-mechanical degrees of freedom
of the
matrix model.

Recently, a large $N$ quantum mechanical model of black hole was proposed
\cite{Chu:2024qil}. The model was motivated by the central idea
of BFSS that matrices can be employed as
the fundamental constituent of quantum gravity.
The use of matrices as space time construct is
also motivated by the fact that it is
extremely natural that quantized space is described by
noncommutative geometry, which becomes
large $N$ matrices when a representation
is taken. We did not impose supersymmetry as we did not start from the
M-theory end. Instead,
the model was constructed bottom-up by requiring the quantum
mechanics to reproduce the defining characteristics of a
quantum black hole in flat (3+1)-dimensions. It was shown that the
Schwarzschild radius and the Bekenstein-Hawking entropy can be precisely
reproduced
by a fuzzy sphere together with a half-filled Fermi sea of states. 
We proposed to identify this as the quantum Schwarzschild black hole.
Our construction  supports the idea that, in quantum gravity,
black holes are properly regarded as highly excited states.
In  \cite{Chu:2024edh},
we showed that the same work for the rotating case and
the Kerr black hole can be
constructed as a rotating fuzzy sphere accompanied by a half-filled
Fermi sea. We give a brief review of the model in section 2.

In this paper we use the model to address the Hawking radiation
  puzzle \cite{Almheiri:2020cfm}.
  What is puzzling is that the semiclassical QFT computation
  gives a final state of Hawking radiation that is thermal.
  This violates the unitarity of
quantum mechanics. 
In the following, we show that Hawking radiation arises in
our quantum mechanical model of black hole via tunneling.
As it turns out, the tunneling is conducted through the
nucleation of a fuzzy monopole. In section 3, 
we construct  the fuzzy monopole as
difference of fuzzy spheres of different ranks.
We show that Dirac equation of fuzzy sphere in a
monopole background admits zero modes determined by the charge
of the monopole. This generalizes the index theorem for the Dirac equation
on a continuum sphere.

Although the fuzzy sphere vacua are stable against perturbative
decay, a fuzzy sphere can tunnel to a smaller fuzzy sphere by
tunneling. 
This is a
  standard problem in quantum mechanics and the tunneling rate can be
  computed from the bounce and the determinant factor once the tunneling
  path is determined. This appears, however, to be a highly
  nontrivial problem in a quantum mechanics with a large number of degrees
  of freedom.  Fortunately this is  where the monopole come to
  save the day. As mentioned above, a crucial property
  of the fuzzy monopole is that they carries 
  fermionic zero modes. As our decay process involves a shrinking
  of the Fermi sea and fermion number is conserved by the Hamiltonian,
  the path integral vanishes unless the tunneling path
  provides fermionic
  zero modes to
  exactly soak up the excess fermi states.
  We show in section 4
  this is exactly what happens for a path conducted by a fuzzy
  monopole.
  We therefore propose that Hawking radiation is described in our quantum
  mechanics as a tunneling process involving the nucleation of a monopole
  from the fuzzy sphere. As the monopole leaves, it also liberates an
  amount of fermi states originally attached to the Fermi sea of
  the fuzzy sphere. We propose to identify
  the released bosonic and fermionic states of the nucleated
  monopole
  with the Hawking radiation. We  show that the
  tunneling rate of the metastable fuzzy sphere agrees precisely
  (up to numerical coefficient)  with the
  semi-classical result of Page for the black hole decay rate
  \cite{Page:1976df}
if the condition \eq{a0} hold. As the condition \eq{a0} is confirmed 
in two independent computations, this is 
a nontrivial test of our model.
Moreover, by focusing on the emission probability  of the
radiation, we find a Boltzmann distribution
with a temperature which can be identified with the Hawking
temperature. In this way a thermal distribution appears to emerge
from the large $N$ system.

We remark on passing that our tunneling is different from the
spacetime tunneling
mechanism of Hawking radiation proposed by
Parikh and Wilczek \cite{Parikh:1999mf}. There, 
Hawking radiation is proposed to arise from the cross-horizon 
passage of particles through classically
forbidden region of spacetime. 
Our analysis offers complementary insights to this spacetime picture
by offering a  microscopic understanding
of the Hawking radiation in terms of quantum mechanical potential
tunneling of the matrix geometry. We note also 
it is amusing that quantum tunneling can occur on a large scale
not just in managed laboratory condition \cite{PhysRevLett.55.1908},
but also for naturally occurring
macroscopic
objects such as black holes.

%

\section{A proposed quantum mechanical model of black hole}

For completeness, we briefly review the proposal \cite{Chu:2024qil} for a
quantum mechanical model of black hole and quantum gravity.
The model of quantum space is given by the
$SU(N)$ quantum mechanics
\be \label{L}
L =  \tr \left[
   \frac{1}{2a_0^2 M_P} \dot{X}^{a2}
   + \frac{M_P}{N^2} \left([X^a,X^b]^2 + 4  X^{a2}\right)
   + i \dot{\psi}^\dag \psi
   - a_2 \frac{M_P}{N^2}\psi^\dag \s^a X^a  \psi \right]
- a_3 r_X M_P .
\ee
Here $X^a_{mn}$, $a=1,2, 3$ represent the coordinates of a
  3-dimensional quantized space,
  $\s^a$ are the Pauli matrices and
  $\psi_{mn\a}$ is a two components spinor, representing a
  fermionic coordinate.
  The model is characterized by three dimensionless coefficients
  $a_0, a_2, a_3$ that parameterize the normalization of the bosonic
  kinetic term, the Yukawa coupling and a possible noncommutative
  background energy term $r_X: = {\rm rank} [X^a, X^b]^2$
  in the quantum mechanics.
  
  The quantum mechanical model admits static solutions
\be\label{FS1}
         [X_a,X_b] = i \e_{abc} X_c,
\ee
which are simply representations of
$SU(2)$. For a spin $j =(K-1)/2$ representation ($K\leq N$) ,
the matrices $X^{(K)}_a$ are $M$ dimensional and describes a fuzzy sphere
$S^2_K$:
\be
X_a^2 = \frac{K^2-1}{4} \id 
\ee
with radius scale like $K$ for $K$ large. 
It was shown in \cite{Chu:2024qil} that 
these  fuzzy sphere solution reproduces the Schwarzschild radius 
  of the (3+1)-dimensional
  Schwarzschild black hole. When the fermion
  is quantized over the fuzzy sphere background, a Fermi sea of states
  is obtained. And it was shown that the counting of microstates for a
  half filled  Fermi sea reproduces precisely the Bekenstein-Hawking
  black hole entropy.
We thus proposed a quantum black hole can be
identified with a fuzzy sphere geometry accompanied with a half-filled
Fermi sea in our model. 
The model has a single mass scale $M_P$ and
  it is related to the Newton constant $G$ as
\be\label{MP}
M_P = \frac{1}{b}\sqrt{\frac{2\hbar}{\pi G}}, \quad 
l_P =  \sqrt{\frac{2\hbar G}{\pi}} ,
\ee
where 
$b := a_2-1+ 2a_3 $ is a constant and the coefficients $a_2, a_3$
are constrained such that $b >0$.  For convenience we have also introduced
a Planck length $l_P$ defined as above.
In addition,  the matching of the angular
momentum of the quantum fuzzy sphere configuration
with that of the Kerr black hole
requires to fix 
$a_0$  to be \cite{Chu:2024edh}
\be \label{a0}
a_0 = \frac{\pi}{3 \hbar}.
\ee
Here we have restored in \eq{MP} and \eq{a0} dependence in $\hbar$.

Classically, an isolated black hole is stable
\cite{Regge:1957td, Whiting:1988vc, Dafermos:2016uzj}. By imposing an
ingoing boundary condition at the horizon and an outgoing boundary
condition at the asymptotic infinity which corresponds to the fact that
no energy is thrown in, one obtain a spectrum of
quasi-normal modes which describe the decay of the perturbations.
In the end, all perturbations are damped except for the $J=0$ and $J=1$
mode. These corresponds
to the mass and angular momentum of the black hole
which are conserved \cite{Israel:1967wq,Israel:1967za,Carter:1971zc}.
In the paper \cite{Chu:2024qil},
linearized perturbation $X^a = X_0^a + \d X^a$ around the
fuzzy sphere $X_0^a$ was considered. At the quadratic level, we get
the  potential for the perturbations, 
\be \label{V2}
V_2 = \frac{2 M_P}{N^2}\tr \d X^a N_{ab} \d X^b,
\ee
where
\be
N_{ab} :=  \left( 
(\ve \cdot L)^2 - (\ve \cdot L) -2 \right)_{ab}
\ee
with $(\ve_a)_{bc} := -i \e_{abc}$ 
being an angular momentum of angular momentum quantum number
$\ell_\ve =1$ 
and $L^a$ is the derivative operator
whose action on a matrix $f = (f_{mn})$ is given by $L^a f = [X_0^a,f]$.
The operator $N_{ab}$ 
is diagonalized by the vector harmonics
$\hat{Y}^{JM_J}_{\ell\; a}$ whose form
can be found in \cite{Chu:2024qil} and will not be
needed here. It was found that the two perturbation modes $\d X^a$
($\ell =0, J =1$ and $\ell =1, J=2$) are classically unstable with
the  tree level potential 
\be \label{V2t}
V^{\rm (cl)}_2 = - \frac{4 M_P}{N^2} \tr \d X^2.
\ee
The mode $\ell =0, J =1$ describes a overall
translation, which is ruled out since the fuzzy sphere is traceless.
As for the $\ell =1, J=2$ mode, one can see that
it is actually stabilized at the  the non-linear level by the quartic
interaction term.
Ultimately, to prove the no-hair theorem, we need to
allow fluctuations to leave the horizon. This can be modeled in our model
by embedding the fuzzy sphere as part of a big matrix with
the complementary block describing the environmental degrees of freedom
and allows the perturbation on the fuzzy sphere to leave.
This is an interesting
problem that we will not pursue here. Instead, we are interested in this
paper the quantum instability of the fuzzy spheres. We
will show that fuzzy sphere is unstable against quantum tunneling.
As it will be clear below, tunneling occurs
entirely due to the physics of the fuzzy sphere, i.e.
the horizon, and this is  why the process can be studied
without introducing the environmental degrees of freedom.

Finally we make a remark on our choice of the mass term. 
It is well known that
for the purpose of having fuzzy sphere as solution in a Yang-Mills type
matrix model,
it is possible to utilize both a tachyonic mass term or
a Chern-Simons mass term \cite{Iso:2001mg}.
Our choice was motivated by the observation 
that the quantum mechanics of the inverted harmonic oscillator (IHO) is
known to capture aspects of the quantum physics of  black holes
\cite{Subramanyan:2020fmx}.  We incorporate a
tachyonic mass term in our model. Our intuition is that the rapid
runaway induced by the tachyonic potential provides a 
quantum-mechanical way to capture the instability resulting from the
large near-horizon redshift
in classical general relativity.
It is possible that,
once the effect of interaction is included,
this mass term may also be responsible for
the appearance of quantum chaos in our model,
a property which has been conjectured
for a quantum black hole \cite{Sekino:2008he,Maldacena:2015waa}.

\section{Monopole on fuzzy sphere}
\label{monopole}

A Dirac monopole on a sphere $S^2$ is a $U(1)$ connection with a
nonzero magnetic flux through the sphere. For nonzero charge $g$, a
description in terms of two patches is needed and the monopole
connection is given
by
\be \label{dirac-A}
A^{(N)} = g (1-\cos \th) d \phi, \quad A^{(S)} = -g (1+\cos \th) d \phi, 
\ee
where $N/S$ denotes the north and the south patches
and $(\th, \phi)$ are the polar
angles on the sphere. The  field strength 
\be
F = g \sin \th d \th d \phi
\ee
is globally defined and gives a flux
\be
\int_{S^2} F = 4 \pi g,
\ee
confirming that $g$ is the monopole charge.
Topologically, the $U(1)$ bundle is
classified by the first Chern number and it  is quantized to be integers:
\be
n:= \frac{1}{2\pi} \int F \in \ZZ.
\ee

In the continuum limit, one may also define a monopole as a gauge field
that twist the
line bundle of matter fields nontrivially, or as a gauge field
which admits Dirac zero modes.
For the last definition, recall that for the monopole
field \eq{dirac-A} of charge $g =n/2$, the
Dirac equation on the sphere,
\be \label{dirac-eom}
\s^a D_a \zeta = 0
\ee
can be written conveniently using the stereographic
coordinates $(z,\zb)$
of the sphere and solved in terms of (anti-)holomorphic functions.
For $n>0$, the local solution near the northern patch is  given by
\be
\zeta =
\begin{pmatrix}
  \frac{P(z)}{(1+ |z|^2)^{\frac{n-1}{2}}} \\0
\end{pmatrix},
\ee
where $P$ is an arbitrary function holomorphic 
in the complex coordinates $z, \zb$.
In general, $\zeta$ blows up near the south pole $z=\infty$,
it can however be extended to a global solution on $S^2$ if $P$ is
restricted to be a polynomial of degree less than or equal to $n-1$.
This gives precisely $n$ zero modes. This is of course guaranteed by the
index theorem. Note that all three definitions of monopole: 
the presence of a nontrivial Chern class,
a nontrivial transition function,  or zero modes, are equivalent
for the case of a continuum sphere.

In the noncommutative case,
topologically nontrivial module on the fuzzy sphere
$S^2_N$ was first constructed in \cite{Grosse:1995jt}. In the large
$N$ commutative limit,
they become 
nontrivial field configurations 
on $S^2$ with a Dirac monopole. The
construction of these nontrivial modules
can thus be taken as a signal for the presence of a
Dirac monopole for finite $N$.
The gauge configuration was however not constructed until
the paper  \cite{Steinacker:2003sd} wherein 
the fuzzy monopole  was constructed as the difference of fuzzy
spheres of different sizes. 
Specifically, it was shown that the matrix connection reproduces
the Dirac monopole connection \eq{dirac-A} in the large $N$ limit.
In the following, we show that this  definition of monopole
also has the nice property of producing the desired number of
fermionic zero modes
for the Dirac equation, even for finite $N$.

The equation of motion of our model is given by
\be \label{eom-A}
   [[X_a,X_b],X_a] + 2 X_b =0.
\ee
A solution of \eq{eom-A} is given by the fuzzy sphere $S_N^2$, which is
a $N$-dimensional representation of $SU(2)$. For convenience, we will
denote the solution  as $X_a^{(N)}$.
Note that $X_a^{(N)}$ act as a
derivative (angular momentum) on
modules on the fuzzy sphere.
In general, given an algebra $\cA$. A
left-$\cA$-module is a vector space $\cV$ together
with a multiplication by
$\cA$ such that $f v \in \cV$ for any  $f\in \cA, v\in \cV$.
A derivative on $\cV$ is defined such that the Leibniz rule holds:
$D_a (f \zeta) = (D_a f) \zeta + f D_a \zeta$.
For example, for the fuzzy sphere $S_N^2$,
$D_a \zeta = X_a^{(N)} \zeta$ for fundamental field $\zeta = (\zeta_m)$
and  $D_a f = [X_a^{(N)}, f]$ for adjoint $f = (f_{mn})$.
We want to add a gauge field $A$ and
consider the covariant derivative 
\be
X_a = X^{(N)}_a - A_a. 
\ee
A particular class of solutions of \eq{eom-A}
is given by representations $X_a^{(K)}$ of $SU(2)$ which describe
smaller  ($K <N$) fuzzy spheres  $S_{K}^2$.
As a result, we obtain the  gauge field on $S_N^2$
\be \label{mono-A}
\An_a :=  \XN_a -  X^{(K)}_a, \quad N>K,
\ee
where $n := N-K$. 
We propose to call \eq{mono-A}  a Dirac monopole on $S_N^2$ with charge
$g=n/2$. To see the reason, let us consider
the zero mode counting for the Dirac equation
on $S_N^2$ for fundamental fermion,
\be \label{dirac}
\s^a D_a \zeta =0.
\ee
Here  $\zeta = (\zeta_m)_{1 \leq m \leq N}$ and 
$\zeta_{m\a}$ is a 2-components spinor, $\a =1,2$. 
To warm up, let us consider first the case with
no gauge field. In this case, $\s^a D_a = \s^a X_a^{(N)}$ is
of full rank and invertible, hence there is no nontrivial zero mode
to \eq{dirac}.
Next, consider $D_a = X_a^{(K)}$ for the gauge field \eq{mono-A}, 
the matrix $\s^a X_a^{(K)}$ is of rank $2K$ 
and so \eq{dirac} admits  $2(N-K)$ nontrivial zero modes
lying in the
kernel of  $\s^a X_a^{(K)}$.
In terms of 2 component spinors, there
are precisely $n = N-K$ such zero modes. This justifies to call
the gauge field \eq{mono-A}  a noncommutative monopole of charge
$n/2$ on the fuzzy sphere $S_N^2$. 

Our proposal can be further justified by considering
the large $N$ limit. Up to irrelevant coefficients,
one can show in  similar fashion to
\cite{Steinacker:2003sd} that the
gauge field reduces to the classical Dirac monopole \eq{dirac-A}
on $S^2$ in the large $N$ limit. Let us consider the explicit
representation of
$SU(2)$. For spin $(K-1)/2$,  $K\leq N$, we have
\bea \label{X-op}
&&(X^{(K)}_3)_{kl} = \d_{kl} \frac{K+1-2k}{2}, \nn\\
&&(X^{(K)}_-)_{kl} = \d_{k, l-1}\sqrt{k(K-k)}, \quad 1\leq k,l \leq K
\eea
where $X_\pm = X_1 \pm i X_2$ and $[X_3,X_\pm] = \pm X_\pm$,
$[X_+, X_-] = 2 X_3$.
As a result, we obtain for \eq{mono-A} the following non-vanishing
matrix elements for $A_a$:
\bea 
(A_3)_{kk} &=& \begin{cases}
  \frac{n}{2} \; , & 1\leq k \leq N-n, \\
(X^{(N)}_3)_{kk}\; , & k > N-n.
\end{cases} \nn \\
(A_-)_{k, k+1} &=&  \begin{cases}
  (X^{(N)}_-)_{k, k+1} \left(1- \sqrt{1-\frac{n}{N-k}}\,
  \right), & 1\leq k \leq N-n, \\
  (X^{(N)}_-)_{k, k+1} \; , &  k > N-n. 
\end{cases}
\label{A-elem}
\eea
To see what \eq{A-elem}
correspond to in the large $N$ continuum limit, let us
introduce the operator Cartesian coordinates on the
unit fuzzy sphere
\be
x_a = \frac{2}{N} X_a^{(N)}.
\ee
In the patch covered by $1\leq k \leq N-n$, which represents the sphere without
the south pole, we obtain in the large $N$ limit the matrix elements
\bea \label{A-large}
A_3 = \frac{n}{2} \id, \quad
A_+ = \frac{n}{2} \frac{x_+}{1+x_3}, \quad
A_- = \frac{n}{2} \frac{x_-}{1+x_3},
\eea
up to a pure gauge transformation in $x_3$.
We may project this to the fuzzy sphere by considering the tangential/projected
gauge
field defined by ${\bf \cA} : = \bA \times \hat{\bx}$. This gives
\be \label{A-tang}
\cA := \cA_a dx_a = \frac{n}{2} (1- \cos \th) d \vphi,
\ee
where we have introduced the polar coordinates of $S^2$:
$x_1 = \sin \th \cos \vphi, x_2 = \sin \th \sin \vphi, x_3 = \cos \th$.
This \eq{A-tang} is precisely the Dirac monopole \eq{dirac-A} 
on $S^2$ with the first Chern number $n$.
We also note that in the large $N$ limit
\eq{A-large} satisfies
\be \label{AgA1}
A^{(n)} = n A^{(1)},
\ee
where the superscript $(n)$ refers to the first Chern number
of the monopole.

\section{Decay of quantum black hole from tunneling of fuzzy spheres}

A stationary black hole emits radiation because  quantum
fields near the horizon produces outgoing flux at infinity.
Semi-classical analysis of quantum field near the horizon give rises to a
thermal spectrum with the Hawking temperature
\be \label{TH}
T_H = \frac{\hbar c^3}{8\pi GM k_B}
\ee
and a decay rate for the black hole \cite{Page:1976df}
\be \label{sc-rate}
\G :=
  -\frac{1}{M} \frac{dM}{dt} = -\frac{Q}{G^2 M^3},
  \ee
where $Q$ is a 
coefficient that depends on the type of emission particles but not on the
black hole itself. 
For example, for pure photon emission,
$Q= \hbar c^4/(15360 \pi)$.
Note that these results are determined in the leading
order of $\hbar$. 
Note also that the  $1/M^3$ dependence is universal for a semi-classical
black hole and is independent of the type of quantum field being
considered. Higher order corrections to these results can be
considered with different source of origin
(e.g. stringy corrections, generalized uncertainty principle, loop
quantum gravity effects etc) and are model dependent.
In \cite{Chu:2024qil,Chu:2024edh},
we have constructed quantum black hole as a particular
configuration of quantum space time. We will now show from
first principle quantum mechanics  that the fuzzy sphere is metastable
against tunneling and the tunneling rate reproduces the $1/M^3$ dependence
for a semi-classical black hole in the leading order of
large $N$ and in the leading
order of $\hbar$. Moreover, 
we will
argue that the Hawking temperature is also reproduced.

\subsection{Tunneling rate in quantum mechanics}

Tunneling in quantum mechanics leads to decay of a metastable vacuum.
In the WKB approximation,  the probability rate may be estimated as
$\G \approx \Omega_{a} |\psi_{\rm WKB}|^2$
where the factor $\Omega_{a}$ represents some kind of attempt frequency
measuring
the rate at which a particle hits the potential barrier
from within a confined structure.  A more accurate and reliable
derivation of the tunneling rate is to use the path integral. 
For a quantum mechanics
\be
L  = \frac{1}{2} m \dot{x}^2 - V(x)
\ee
with a false vacuum at $x_f$,
the tunneling decay rate of the false vacuum
can be extracted from the path integral of the retention amplitude
\cite{Coleman:1985rnk,Devoto:2022qen}
\be
\la x_f | e^{-HT} |x_f \ra = \int_{x(-T/2) = x_f}^{x(T/2)=x_f} Dx\; e^{-S/\hbar}
\ee
in the limit of large $T$. Summing over the bounce,
one obtain the decay rate
\be \label{rate}
\G = K e^{-S_b/\hbar},
\ee
where
\be \label{Sb}
S_b = 2 \int_{x_f}^{x_t} dx \sqrt{2m(V(x)-V(x_f))}
\ee
is the bounce action and 
\be \label{K}
K = \frac{1}{2}
\sqrt{\frac{S_b - S_f}{2\pi \hbar m}}
\sqrt{\frac{{\rm Det} S''[x_f]}{{\rm Det}' S''[x_b]}}
\ee
is a determinant factor
which capture the path integral
contribution from the quantum fluctuations around the bounce.
Here $x_t$ is the location of the turning point
of the potential, $ S_f = S[x_f]$, $S_b = S[x_b]$
and $x_b(\t)$ is the bounce trajectory. 
The bounce satisfies the equation of motion (in  Euclidean time),
\be
\frac{1}{2} m \dot{x}^2 - V(x) = -V(x_f)
\ee
and the boundary condition
\be
x(-T/2) = x(T/2) =  x_f, \quad \dot{x}(0)=0,
\ee
which expresses the fact that the bounce
starts initially at $x_f$, reaches the turning point $x_t$
at $\t =0$ and returns to $x_f$ at $\t=T/2$. As a result, one obtains
the action \eq{Sb} for the bounce. This reproduces immediately 
the usual WKB transmission probability:
$e^{-S_b}= |\psi_{\rm WKB}|^2$.
The tunneling rate is given by the transmission probability times the
factor $K$ which capture  the bouncing rate at which the particle hits
the potential barrier. 
The determinant factor can be evaluated
using the Gelfand-Yaglom formula for
functional determinant.
The result  is \cite{Devoto:2022qen}
\be \label{K1}
K =\Omega_0^{3/2} m^{1/2} \hat{a}\; \frac{1}{\sqrt{\pi \hbar}}.
\ee
Here $\Omega_0$ is the frequency of oscillation at the false vacuum,
and 
$\hat{a}$ is a shape factor that depends on the shape of the
potential in the region between the false vacuum and the turning point:
\be\label{ahat}
\hat{a} = \lim_{\e \to 0} \e \exp \left[m \Omega_0 \int_{x_t}^{x_f -\e}
  \frac{dx}{\sqrt{2m (V(x)-V(x_f))}}\right].
\ee
We note that the tunneling rate \eq{rate}
is for a bosonic quantum mechanics. When there are
fermions (Grassmanian variables),  the path integral imposes selection
rule on the transition amplitude due to presence of fermionic
zero modes. We will discuss this important point next.

\subsection{Zero modes analysis and monopole channel}

We have proposed that the fuzzy sphere solutions in our quantum
mechanics together with a half-filled Fermi sea of states describes a
quantum black hole. For a fuzzy sphere of rank $N' = N-n$ with $n \in \NN$
fixed and finite, it has energy
\be
E^{(N')} = \frac{bN' M_P}{2}
\ee
in the leading order of large $N$.
Although the
fuzzy sphere vacua are locally stable, tunneling
may occur from a fuzzy sphere of a higher rank $N$ to one with a
lower rank $N'$. This is accompanied with a
reduction in the Fermi sea of states.
Since the Hamiltonian of the theory conserves fermion
number, tunneling is possible only if the decay path admits the
right number of zero modes to soak up the excess  fermionic
states.
In the following, we show that a tunneling path described by the
emission of a monopole satisfies the selection rule precisely. And this
leads
to a decay rate that scales with the black hole mass $M$
exactly as \eq{sc-rate}. As a result,
Hawking radiation of black hole is
contained in our quantum mechanical model.

In our matrix quantum
mechanics, the decay rate of the metastable state can be exacted from the
path integral of the retention amplitude
\be \label{RA}
\la X_f, \psi_f | e^{-HT} |X_f,\psi_f \ra = \int_{X(-T/2) =X_f}^{X(T/2)=X_f}
[DX] \; e^{-S_{\rm bos}[X]}
\int_{\psi(-T/2) = \psi_f}^{\psi(T/2) = \psi_f} [D \psi  D \psi^\dag] \;
e^{-\int \tr \psi^\dag (\del_\t - c \s\cdot X  )    \psi} 
\ee
Here $X_f = \XN$ is the false vacuum, $\psi_f$ is the accompanied
half-filled Fermi sea with $N^2$ states,
$c = a_2 M_P/N^2$ is an unimportant numerical coefficient and 
\be
S_{\rm bos} = \int d\t\;  \tr \left(
\frac{1}{2a_0^2 M_P} \dot{X}_a^{2} +V(X) \right)
\ee
is the (Euclidean) bosonic action with
\be \label{V}
V(X) = 
   - \frac{M_P}{N^2} \left([X_a,X_b]^2 + 4  X_a^2\right)
   \ee
the classical potential.
Summing over
the contribution of the classical bounce solution $X=X_b(\t)$
gives the saddle point approximation for the decay rate,
\be \label{KF}
\G = 
K e^{-S_b/\hbar} \times F.
\ee
Here $S_b$ is the bounce action, $K$ is the bosonic determinant around
the saddle point and 
\be \label{F}
F := \int [D \psi D \psi^\dag] \;
e^{-\int \tr \psi^\dag (\del_\t - c \s\cdot X_b  ) \psi}  
\ee
is the fermionic determinant around the saddle point.
In general, the bounce admits femionic zero modes  defined by the equation
\be \label{zm-adj}
(\del_\t -c \s \cdot X_b(\t)) \psi_0 =0, \quad
 -T/2 \leq \t \leq T/2, 
 \ee
and the boundary condition $\dot{\psi_0}(0) =0$, which
is the counterpart of the bosonic boundary condition $\dot X_b(0) =0$.
As a result, $F$ factorizes to 
\be
F = F_0 \times {\rm Det}'[\del_\t - c \s\cdot X_b],
\ee
where
\be
F_0 = \int [D \psi_0 D \psi_0^\dag]
\ee
gives the zero mode contribution.
We note that as $\psi_{0 kl \a}$ 
can be considered a fundamental fermion
with a flavor index, the solution of \eq{zm-adj} can be expanded as
\be
\psi_{0 kl \a} = \sum_{i=1}^{n_Z} c^{(i)}_l
\z^{(i)}_{k\a} + \ct^{(i)}_l \z^{(i) \dag}_{k\a}
\ee
where $c^{(i)}_l, \ct^{(i)}_l$ are Grassmanian coefficients and
$\z^{(i)}_{k\a}$, $i = 1 \cdots, n_Z$ is a basis of  solution
of the zero mode equation
\be \label{zm-fund}
(\del_\t -c \s \cdot X_b(\t)) \zeta =0
\ee
in the fundamental representation.
As a result, the zero mode factor is a $2Nn_Z$ dimensional integral
\be
F_0 = \int \prod_{i,l} Dc^{(i)}_l  D\ct^{(i)}_l.
\ee
The presence of  nontrivial zero modes $F_0$ imposes
 a stringent constraint on the possible decay paths.
 
The computation of the tunneling rate hinges on the choice of the
decay path.  In principle one should consider all possible saddles and
select the one that produces the highest tunneling rate.  This gives
the dominant decay channel and the corresponding path is called the
MPEP (most probable escape path) in \cite{Banks:1973ps}.
This is a difficult
task for our large $N$ quantum mechanics and we need some good
intuition in reducing the degrees of
freedom and  guessing the relevant  path. 
We start with the observation that it has been observed empirically 
\cite{Banks:1973ps}
that the MPEP
is often given by a straight line path.
This suggests for us to consider the following interpolating path
\be \label{path1}
X = (1-\gt) \XN - \gt \XNp, \quad 0 \leq \gt \leq 1 
\ee
for the decay $\XN \to \XNp$ in our quantum mechanics.
A saddle $\gt =\gt(\t)$  with the boundary condition
$\gt (-T/2) = \gt(T/2) =0,  \gt(1) =0$ describes a bounce
for the decay process $\XN \to \XNp$. Note that \eq{path1} may
be rewritten as
\be\label{path2}
X = \XN - \gt \An,
\ee
where
$\An = \XN -\XNp$ is a fuzzy monopole with the first Chern number
$n= N-N'$. In this form, the path
describes a tunneling process that involves 
the nucleation of a monopole from the original fuzzy sphere. 
Although the possibility of understanding the quantum  mechanical
decay of black hole spacetime as
a tunneling process involving the nucleation of a monopole is quite
appealing, a priori it is not obvious that it can work. As explained
above, transitions between fuzzy spheres
with different fermion number are forbidden unless the
bounce itself provides the right number of fermionic zero modes to
saturate the excess of fermi
states in  the initial wavefunction. 

In fact, if we split the bounce into two parts $-T/2 \leq \t \leq 0$
and $0\leq \t \leq T/2$, then tunneling is possible only if it is
possible to transit from $-T/2$ to 0 and then back to $T/2$.  Now at
$\t =-T/2$ we have a set of $N^2$ fermionic states in the Fermi sea of
the original fuzzy sphere vacuum $\XN$, and at $\t =0$ we have a set of
$N'{}^2$ fermionic states unitary related to the Fermi sea of the
vacuum $\XNp$.  As a
result, the transition is allowed only if the excess $\D N_F = 2 Nn$
(in the leading order of large $N$) of fermi states in the original
vacuum is precisely saturated by the fermionic zero modes. This gives
$n_Z =n$, meaning that the bouncing path for the decay from $\XN$ to
$\XNp$ must have precisely $n = N-N'$ zero modes.  Quite amazingly,
this is exactly achieved by the monopole saddle path \eq{path2},
To see this, we note that as the zero mode equation \eq{zm-fund}
is of first order,
its solution is fixed entirely by the value of $\zeta$ at some point.
It is convenient to consider the point $\t=0$ since due to the
bounce boundary condition $\dot{\z}(0) =0$. Using the fact that
$X_b (0) = \XN - \An$, we arrive exactly
at the monopole equation \eq{dirac}. This gives
$n$ zero modes which is precisely what we need. We remark that
the path \eq{path2}
is not only sufficient, it is reasonable to believe that it is also the
only possible path that has the desired zero modes structure and hence
allows a nonvanishing transition amplitude.

\subsection{The bounce and the decay rate}

Let us now proceed to determine the barrier potential.
For this purpose, it is convenient to consider
a parameterization of the path \eq{path1} from the midpoint. 
Introducing $-1\leq \g := 2 \gt -1 \leq 1$, we have
\be \label{path3}
X = X^* - \frac{\g}{2} \An, \qquad
X^* := \frac{1}{2}(\XN+ \XNp).
\ee
Here $X^*$ denotes the mid-point of the two fuzzy spheres and 
$\g =-1$ (resp. 1) corresponds to the fuzzy sphere $\XN$ (resp. $\XNp$).
Note that for even $n$, the monopole charge $g=n/2$ is an integer.
In the large $N$ limit, the midpoint can be approximated  by the
fuzzy sphere
\be \label{X*Xg}
X^* = X^{(N-g)}
\ee
and \eq{path3} becomes
\be \label{path4}
X = X^{(N-g)} - \frac{\g}{2} \An.
\ee
Using also  
\be
A^{(n)} = 2g A
\ee
where $A:=A^{(1)}$ denotes the monopole potential \eq{A-elem}
with unit first Chern number. Therefore, 
we obtain for integer $g$ the barrier potential
$V(\g) := V(X(\g))$,
\be \label{Vg}
V(\g) = - \frac{2 g^2 M_P}{N^2}\tr A^2\;  \g^2
-\frac{g^3 M_P}{4N^2}  \g^3
 + \frac{g^4 M_P}{96 N^3} \g^4.
 \ee
Note that there is no linear term in $\g$ as $X^{(N-g)}$ is a saddle point
 of $V$.  To obtain \eq{Vg}, we have used the following
 trace formula for the matrix elements of the monopole 
 \be
 \tr ([A_a,A_b][\XN_a,A_b]) = -\frac{1}{16} + O(\frac{1}{N}),
\qquad
 \tr ([A_a,A_b]^2) = -\frac{1}{96N} + O(\frac{1}{N^2}).
 \ee
 We have also ignored a constant term $-NM_P/2$ in
 the expression above.
 It is interesting to note  that
 \be \label{trA2}
\tr A^2 = \frac{N \log N}{4} + O(1), 
 \ee
has a non-analytic $\log$-dependence instead of a
polynomial dependence in $N$.
This gives rises to a $\log N$ dependence in the large $N$ expression
for the potential.
This point is important as we will show now that the bounce action
$S_b$ inherits such a dependence and results in a decay
probability that is suppressed by powers of $1/N$ instead of an
exponential suppression.   Without
this it would be impossible to match up with the semi-classical result
of black hole decay rate \eq{sc-rate}. Since
 the $\g^3$ and $\g^4$ terms can be ignored for
 large $N$, we obtain the  potential
 \be \label{Vg2}
V(\g) = - \frac{2g^2 M_P }{N^2} \tr A^2 \;   \g^2 + O(\frac{1}{N^2}). 
\ee
in the leading order of large $N$.

\begin{figure}
                        \centering
                        \includegraphics[scale=0.4]{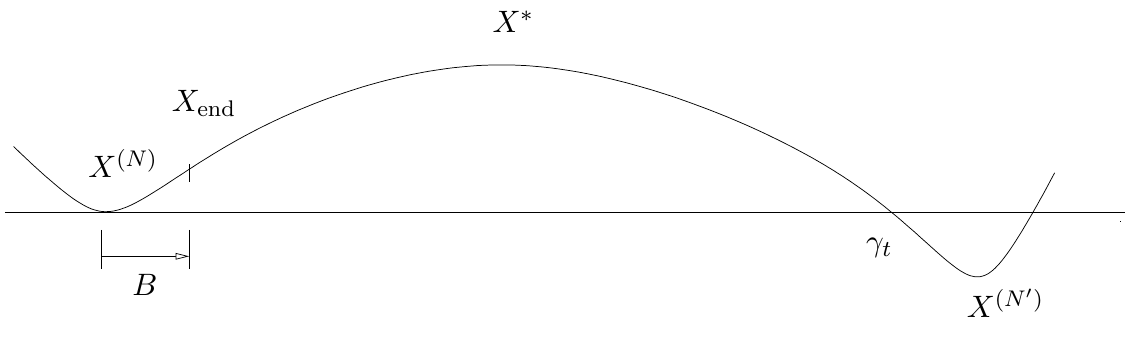}
                        \caption{Potential barrier for the monopole
                          channel between two fuzzy spheres}
                        \label{pot}
\end{figure}

Note that
the potential admits a maximum at $\g=0$ and represents a barrier.
We also note that due to the approximation we used, the potential is not
able to cover the region near the fuzzy sphere vacuum $\XN$. To see this,
let us consider 
a small fluctuation $\d X = -\gt \An$ along the path \eq{path2}
 near the fuzzy sphere vacuum
 $\XN$. Using the perturbation formula \eq{V2}, and rewriting using
 $\g$, we obtain 
the Lagrangian
\be \label{Lgt}
L = \frac{n^2}{4}\tr A^2  \left(
\frac{\dot{\g}^2}{2 a_0^2 M_P} - \frac{8 M_P}{N^2} (\g+1)^2
\right), \quad \g \approx -1.
\ee
This shows that the the fuzzy sphere
$\XN$ describes a local minimum of the potential
with a vacuum oscillation frequency
\be \label{om0}
\Omega_0 = \frac{4a_0M_P}{N}.
\ee
It is however clear  that \eq{Vg2} does not produce such a minimum.
The reason why \eq{Vg2} fails to do so is because we have
 approximated the original path \eq{path3} by \eq{path4} in order to get
 \eq{Vg2}. 
 While \eq{X*Xg} is true point-wise (i.e. for each matrix element)
 in the large $N$ limit, it is easy to show that the distance between
 the two points is in fact finite, i.e.   $\tr (\D X^2) = O(1)$ where
 \be
\D X_a := X_a^* - X_a^{(N-g)}.
\ee
This implies that the endpoint of the path \eq{path4},
\be \label{DX}
X^{\rm (end)} := X(\g =-1) = \XN + \D X,
\ee
is also separated from $\XN$ by the same distance, i.e. the path
\eq{path4} actually ends too early. This explain why \eq{Vg2}
does not include the local behavior \eq{Lgt} near $\XN$. To obtain
the full barrier potential, we can
patch up \eq{Vg2} with the local potential \eq{Lgt}:
\be \label{matching-V}
V = \begin{cases}
  \frac{1}{2}m \Omega_0^2
  (\g +1)^2), & |\g +1| \lesssim B, \\
  \frac{1}{2}m  \Omega^2 (\g_E^2 - \g^2)
  + \frac{1}{2}m \Omega_0^2 B^2, & -\g_E \leq \g \leq  \g_t ,
\end{cases}
\ee
where  $B$ specify the region around $\XN$ where we can use the
description \eq{Lgt}, $\g_E \lessapprox 1$ denotes the matching point,
and we have defined, for convenience,
\be \label{m-omega}
m = \frac{g^2}{a_0^2 M_P} \tr A^2, \quad
\Omega = \frac{2 a_0 M_P}{N} = \frac{1}{2} \Omega_0.
\ee
See figure \ref{pot} for  the potential.


To implement the matching,
$B$ has to satisfy
\be
\frac{1}{2}m \Omega_0^2 B^2 \sim \D E
\ee
where 
\be
\D E : = V(X^{(\rm end)}) - V(\XN) 
\ee
is the potential energy difference
between $\XN$ and $X^{(\rm end)}$.  With the perturbation \eq{DX}, it is
easy to use \eq{V2} to determine $\D E$. We obtain 
\be
\D E \sim M_P/N^2 .
\ee
As a result,
\be \label{Bm}
B \sim \frac{1}{a_0\sqrt{m}} \sim  \frac{1}{\sqrt{N \log N}}.
\ee
This result will be useful later.

Next let us determine the bounce action. Using \eq{Vg2} and rotated
to Euclidean time, the bounce action is given by $S_b = 2 S_{\rm WKB}$
where
\be
S_{\rm WKB} = \int_{-1}^{\g_t}
d \g \sqrt{2m( V (\g) - V(-1))}.
\ee
One can see that the contributions from the regions
$(-1, -\g_E), (\g_E, \g_t)$ are higher order in $1/N$ and can be ignored.
As a result, we obtain
\be
S_{\rm WKB} = m \Omega C (\th_E+ \frac{1}{2} \sin 2 \th_E)
\ee
where $C^2 = \g_E^2 + 4 B^2$, and $\th_E$ is determined by
$ \sin \th_E = \g_E/C$. Since $B$ is small, we have
$\th_E = \pi /2$ and
\be
S_b = \frac{g^2 \pi }{2 a_0} \log N
\ee
in the leading order of large $N$. 
Now
in the analysis \cite{Chu:2024edh}
of the Kerr black hole, the constant $\a_0$ has
been determined to be 
\be
a_0 = \frac{\pi}{3\hbar}
\ee
by a  matching of the angular momentum.
Therefore,
\be
S_b =\frac{3g^2 \hbar}{2} \log N.
\ee
The leading contribution is given by the charge 1 process and we obtain
the contribution
\be \label{eS}
e^{-S_b/\hbar} \sim \frac{1}{N^{3/2}}, \quad \mbox{for $N \to N-2$}
\ee
to the decay rate of the fuzzy sphere. The nucleation of monopole
with $g>1$ gives a contribution of $1/N^{3g^2/2}$ and is of higher order
in $1/N$. These 
are negligible for $N$ large, but  can become significant for the decay
of small black hole.

Finally, we look at the determinant factor $K$ as given by \eq{K1}.
Note that the quantities $\Omega_0$ and $m$
are local properties of  the vacuum $\XN$, while $\hat{a}$
is a shape factor that depends on the shape of the barrier potential 
\eq{Vg}. With the potential \eq{matching-V}, one can split the integral
\eq{ahat} into two parts. The monopole potential contributes a
factor that is independent of $N$, while the part 
around the fuzzy vacuum contributes a factor of $B$. Therefore, up to an
$N$ independent overall constant
\be
\hat{a} = B.
\ee
Substituting \eq{m-omega} and \eq{Bm} into \eq{K1}, we obtain 
\be
K = \frac{1}{l_P N^{3/2}}
\ee
up to a dimensionless numerical constant.
The quantum mechanical decay rate is then given by 
\be
\G \sim \frac{1}{l_P N^3} \sim \frac{\hbar}{G^2M^3}.
\ee
This reproduces, 
up to a numerical constant, the
semi-classical decay rate \eq{sc-rate}  of black hole.

\subsection{Hawking radiation and temperature}

\begin{figure}
  \centering
\includegraphics[scale=0.6]{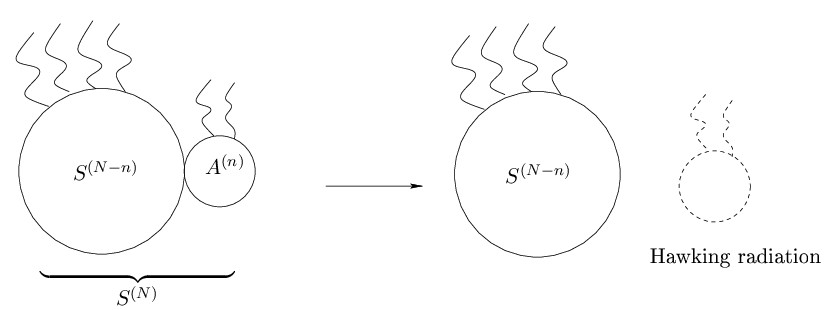}
  
                     \caption{Monopole and Hawking radiation}
                        \label{hawk}
\end{figure}

In the above we have used the path integral formulation
to compute the decay rate for the tunneling process
\be
\Psi (t=-T/2, \XN, {\rm FS}) \longrightarrow \Psi(t=T/2, \XNp,{\rm FS'}),
\ee
where ${\rm FS}$ (resp. ${\rm FS'}$)
stands for the half filled Fermi sea of the vacuum
$\XN$ (resp. $\XNp$). In the path integral description,
the excess amount of fermionic
states in the initial vacuum
is saturated by the zero modes of the monopole.
Physically, as the monopole nucleates from $\XN$ to
leave behind $\XNp$, 
the excess fermi states also escapes from $\XN$.
We propose to identify
the released  quanta of the quantum space with the
Hawking  radiation.
We note that in the path integral description,
one specify the initial and the final states, and the details
of the intermediate state is not available.
On the contrary, in a real time formulation of the tunneling problem,
one should be able to follow the time evolution of the total
wave function and study not just the decay of the fuzzy sphere
(black hole), but also
the emitted
Hawking radiation in a fundamental quantum
mechanical way.

Without launching on to the full computation, we can amend the
path integral computation with a probability description of the emitted
quanta. Ignoring the small region of size $B$ near $\XN$,
the Lagrangian for our tunneling process is given by
\be
L = \frac{1}{2}m \Omega^2 (\dot{\g}^2 + \g^2) + {\rm constant}.
\ee
Suppose at an intermediate position $\g_1$, the system emits a quanta of
energy $\o$ and the configuration get back reacted to reach
the point $\g_2$. By energy conservation, we have
\be
E_1= E_2 + \o, \quad \mbox{where} \quad
E_i:= \frac{1}{2}m \Omega^2 \g_i^2
\ee
Let us denote the probability of emitting a
bit of quantum space of energy $\o$
by $P(\o)$.
Now, whether the configuration
of energy $E_1$ decays directly to the final fuzzy sphere configuration
or instead first emits a quantum and subsequently the remaining system
decays to the fuzzy sphere, the final product is the same, and
consistency requires that the transition
rate be the same for both descriptions.
We obtain
\be
\G_{E_1} = P(\o) \G_{E_2}. 
\ee
It is easy to obtain that
\be
e^{-S_b}|_{E_i} = \left(\frac{1}{N^{3/2}}\right)^{\g_i^2}.
\ee
As a result, we obtain 
\be \label{PT}
P(\o) = e^{-\o/T}, \quad \mbox{where} \quad T := \frac{\Omega}{2\pi}.
\ee
This is a Boltzmann distribution for a thermal system of temperature
$T$.
One can express $T$ in terms of $N$ 
\be
T = \frac{M_P}{3N}.
\ee
It is nontrivial that
it has exactly 
the same $N$-dependence as the Hawking temperature \eq{TH}
\be
T_H = \frac{b M_P}{8N}.
\ee
This is nontrivial.
One can reproduce the Hawking temperature if one
chooses the parameters in the model to have
\be
b = 8/3. 
\ee
It is interesting that a thermal probabilistic distribution
arises in our fully quantum mechanical system of a large $N$ system.
However, we emphasis that the Hawking radiation is not a pure thermal state
in our model. In  a real time formulation for the tunneling problem,
one will be able to determine fully
the wavefunction of the Hawking radiation
and  recover, beyond the
probabilistic thermal description,
the quantum information contained therein.

\section{Discussions}

In this paper, we have provided further justifications to the model of
quantum black hole proposed in \cite{Chu:2024qil} by showing that
Hawking radiation has its origin as tunneling of fuzzy sphere system
in our model.
We remark that in the usual analysis of Hawking radiation, the Hawking
particles arises from the quantizing the fields of
the standard model in the curved spacetime. So 
the Hawking temperature found there belongs to the the matter fields,
and strictly speaking, not necessary a property of the spacetime. 
This is different in our model where
the emitted Hawking quanta  are part of
the quantum spacetime that makes up the black hole.
So  the temperature we found can be more properly
considered as a temperature
of the quantum spacetime itself. We remark also that the
Hawking radiation found here arises from the
quantum space that are originally making up the black hole. 
This is amazingly close to the model
proposed in \cite{Chu:2022ieq,Chu:2023mqi}, which was
proposed as a complementary spacetime picture 
of the Page curve originally achieved in
the holographic island proposal \cite{Engelhardt:2014gca,
Penington:2019npb,Almheiri:2019psf, Almheiri:2019hni}. 

The inverted mass term is seen to play a crucial role in our analysis.
We note that the one dimensional (1d)
IHO has been shown to carry an exponential growing
OTOC \cite{Hashimoto:2020xfr} that satisfy the chaos bound
\cite{Maldacena:2015waa}.
Although this OTOC in the simple
1d quantum mechanics is due to  operator growth
and has nothing to do with quantum chaos,   it does
not exclude the possibility that
quantum chaos may appear in our model
once the mixing effect of interaction is taken into account.
The display of chaos and reproduction of
the scrambling time for a quantum black hole
will be another important check of the model.

In the above we have determined the emission probability of Hawking
radiation with a
consistency argument. In principle, one can go beyond and consider the
full wavefunction of the multi-partite state. This wavefunction contains
multi-partite entanglement information which cannot be seen at the level
of probability.  A numerical
stimulation of the finite $N$ quantum mechanics may be a
viable interesting option
\footnote{We thank Jun Nishimura for
stimulating discussion on this.}.
This will allow to  develop observables to describe
how the Hawking radiation departs from thermality.
It has been conjectured \cite{Iizuka:2025ioc}
that the multi-partite entanglement
entropy for Hawking radiation also follows  a Page curve.
It is interesting to check this with our quantum mechanical model. 

Our model also offers a concrete large $N$
model of black hole with fermions
and specific interaction where one may check
the conjecture of memory burden \cite{Dvali:2020wft}
for quantum black hole. 
In this paper, we have consider black hole with macroscopic size
(with $N$ large). It is interesting to study fuzzy spheres with
finite $N$. These are truly quantum mechanical object and do not admit
any classical geometric description. The properties of these
``mini-size black hole'' would  be relevant for the discussion
of primordial black hole and remnants.

Sometime ago, \cite{Almheiri:2012rt} has argued that the following
three statements cannot all be true:
(i) Hawking radiation is in a pure state,
(ii) the information carried by the radiation is emitted from the region
near the horizon, with low energy effective field theory valid beyond
some microscopic distance from the horizon, and
(iii) the infalling observer encounters nothing unusual at the horizon. 
In our model, (i) and (ii) are true but not (iii) since
the classical horizon of black hole is replaced by a fuzzy sphere
of noncommutative geometry. As a co-dimension one object, one can
think of it like a thin shell. However unlike thin shell made of matters e.g.
\cite{Kraus:1994by,
  Cardoso:2016wcr,Kawai:2013mda}, this thin shell is  a crack
of quantum space and behaves like a domain wall with
some energy and momentum.
This suggests a different structure of vacuum across the junction
\cite{Israel:1966rt,Chu:2021uec} and this can be felt by an infalling observer.

According to our model, black hole is made up quantum space composing of
three bosonic and a complex fermionic matrix coordinates.
The fermionic part is not directly seen in the  classical description,
but plays crucial role in accounting for the microstates counting and in the
decay of quantum black hole. It is interesting to investigate further the
other macroscopic physical effects (e.g. astrophysical or particle
phenomenological) of this hidden sector of the black hole.

The Bohr model identified the fundamental constituents of the atom and
the existence of stationary eigenstates — both of which became
foundational to the development of quantum mechanics. It is hoped that
analogous lessons drawn from the quantum black hole model
\cite{Chu:2024qil} , such as the
role of quantum space as the elementary building block of black
hole, may likewise inform the construction of the theory of quantum
gravity. We note  it is possible that our proposed model is for
a single black hole in isolation, with quantum gravitational
interactions not yet included. This situation parallels the Bohr
model, where incorporating atomic interactions ultimately required the
introduction of gauge fields and the full QFT machinery of QED.
Nevertheless, a salient feature of the large $N$
matrix model \cite{Banks:1996vh, Ishibashi:1996xs, Taylor:2001vb}
is that it is already second-quantized in character: multi-body
configurations appear naturally within the description as block
structures. It is therefore quite possible that quantum gravity is
already implicit in the proposed quantum mechanics of quantum
spaces. In that case, a central challenge is to understand how
Einstein gravity emerges in the large $N$ continuum limit\footnote{We
thank Satoshi Iso for discussions on the large $N$ limit of matrix
models and on possible subtleties therein.}. A probe analysis may help
to elucidate geodesic motion and the emergent metric. It would also be
illuminating to understand how holography \cite{Maldacena:1997re,Ryu:2006bv}
arises from the many-body
mechanics of quantum space. The entanglement structure  of the matrix degrees
\cite{Ryu:2006bv,Swingle:2009bg} of
freedom appears to be a fascinating subject of study.

\section*{Acknowledgments}

We thank Harold Steinacker for discussion and
helpful comments on the manuscript, especially the part on monopole.
We also
thank Dimitrios Giataganas, Pei-Ming Ho, Norihiro Iizuka, Satoshi Iso,
Hiroshi Itoyama,
Hikaru Kawai,
Jun Nishimura and Tamiaki Yoneya
for useful discussion.
We acknowledge the support of this work by NCTS, the  
National Science and Technology Council of Taiwan for the
grant 113-2112-M-007-039-MY3, and
the National Tsing Hua University 2025 Talent
Development Fund for a TSAI WANG, YUAN-YANG Distinguished Talent Chair
professorship.

\bibliographystyle{utphys}

\bibliography{references}  
\end{document}